\documentclass[prl,aps,twocolumn,showpacs,amsmath,amssymb]{revtex4}

\usepackage{graphicx}
\usepackage{color}

\usepackage{amsfonts,amsmath}
\usepackage[dvips]{epsfig}
\usepackage[latin1]{inputenc}
\usepackage{bbm}

\newcommand{\rb}{\protect{\bf r}}
\newcommand{\Rb}{\protect{\bf R}}
\newcommand{\kb}{\protect{\bf k}}
\newcommand{\Kb}{\protect{\bf K}}
\newcommand{\Eb}{\protect{\bf E}}
\newcommand{\Pb}{\protect{\bf P}}
\newcommand{\Mb}{\protect{\bf M}}
\newcommand{\Hb}{\protect{\bf H}}
\newcommand{\pb}{\protect{\bf p}}
\newcommand{\Gb}{\mathcal G}
\newcommand{\Gbn}{{\mathcal G}^{(0)}}
\newcommand{\abs}[1]{\protect{\left| #1\right|}}
\newcommand{\eins}{\protect{\mathbbm 1}}

\begin{document}

\title{Local-field effects in radiatively broadened magneto-dielectric
media: negative refraction and absorption reduction}
\author{J\"urgen K\"astel and Michael Fleischhauer}
\affiliation{Fachbereich Physik, Technische Universit\"{a}t
Kaiserslautern, D-67663 Kaiserslautern,
Germany}

\author{Gediminas Juzeli\=unas}
\affiliation{Institute of Theoretical Physics and Astronomy, Vilnius University, A Go\v{s}tauto 12, Vilnius 01108, Lithuania}

\date{\today}

\begin{abstract}
We give a microscopic derivation of the Clausius-Mossotti relations for a homogeneous and isotropic 
magneto-dielectric medium consisting of radiatively broadened atomic oscillators. 
To this end the diagram series of electromagnetic propagators is calculated exactly
for an infinite bi-cubic lattice of dielectric and magnetic dipoles for a lattice constant
small compared to the resonance wavelength $\lambda$. 
Modifications of transition frequencies and linewidth 
of the elementary oscillators are taken into account in a selfconsistent way
by a proper incorporation of the singular self-interaction 
terms.  We show that in radiatively broadened media sufficiently close to the free-space resonance 
the real part of the index of refraction approaches 
the value -2 in the limit of $\rho \lambda^3 \gg 1$, where $\rho$ is the
number density of scatterers. Since at the same time the imaginary part vanishes as $1/\rho$
local field effects can have important consequences for realizing low-loss negative index materials.
\end{abstract}

\maketitle


\section{Introduction}


It is well known that in dense dielectric materials the induced polarization $\Pb$ alters the
field strength $\Eb_{\rm loc}$ acting on the constituents (i.e. the local field)
compared to the average macroscopic field $\Eb_{\rm m}$. Macroscopic considerations
show that  in systems with high symmetry such as
a cubic lattice the two fields are related to each other according to
$\Eb_{\rm loc}=\Eb_{\rm m} + \Pb/(3 \varepsilon_0)$ \cite{Jackson,LL}. This leads to the well-known Clausius-Mossotti relation
for the permittivity $\varepsilon(\omega)$
\begin{equation}
\varepsilon(\omega) = 1 +\frac{\rho\, \alpha(\omega)/\varepsilon_0}{1-\rho\,\alpha(\omega)/(3 \varepsilon_0)}
\label{eq:Clausius-Mossotti-dielectr}
\end{equation}
where $\rho$ is the density and $\alpha(\omega)$ the polarizability of the oscillators.
Similar arguments hold
for a purely magnetic material \cite{Cook}, except that the required densities are usually much higher due to
the smallness of magnetic dipole moments and polarizabilities.
In  linear response $\alpha(\omega)$ is well 
described by a damped-oscillator model \cite{Jackson}
\begin{equation}
\alpha(\omega)=\alpha^\prime + i\, \alpha^{\prime\prime}=\frac{\alpha_0}{\omega_0^2-\omega^2-i\gamma\omega}.
\label{eq:alpha}
\end{equation}
The corresponding (real-valued) parameters such as  the oscillator strength $\alpha_0$, the resonance frequency and width,
$\omega_0$ and $\gamma$, are determined by the microscopic model.
In general the linewidth $\gamma$ contains radiative as well as non-radiative contributions.
For purely radiative interaction these parameters are strongly affected by the renormalization
of energy levels and spontaneous emission processes caused by the interaction with the vacuum electromagnetic field
in the medium \cite{Nienhuis-Physica-1976,Knoester-PRA-1989,Glauber-PRA-1991,Barnett-PRL-1992,Juzeliunas-PRA-1997,Scheel-PRA-1999,Fleischhauer-PRA-1999,Dung-PRA-2003}.
Since the mode structure of the electromagnetic field inside a dense medium can be substantially modified compared to free space,
one would expect that the polarizability entering eq.(\ref{eq:Clausius-Mossotti-dielectr}) is different from that in free space. 
In a macroscopic approach $\alpha(\omega)$ is however an input function and no conclusion can be drawn about 
possible changes due to the different structure of the vacuum modes inside the medium. To take into account the modification of transition
frequencies and radiative linewidth in a dense medium in a self-consistent way requires a microscopic approach.

In the present paper we develop a microscopic approach to local field effects in dense materials with simultaneous
dielectric and magnetic response using Greensfunction techniques similar to those used by deVries and
Lagendijk for purely dielectric materials \cite{Lagendijk}. To this end we consider an infinitely extended bi-cubic
lattice of electric and magnetic point dipoles with isotropic response with a lattice constant small compared to the
transition wavelength. We however do not make use of the assumptions made in \cite{Lagendijk} to
renormalize the singular self-interaction contributions to the lattice  $T$-matrix which eliminated 
radiative contributions to linewidth and transition frequencies altogether. We show that instead the self-interaction 
contributions can be summed to yield the dressed $t$-matrix of an isolated oscillator interacting with the
vacuum modes of the electromagnetic field in free space. In this way we derive Clausius-Mossotti relations
for general, radiatively broadened, isotropic magneto-dielectrica. 
Apart from non-radiative broadenings,
the electric and magnetic polarizabilities entering 
these equations are shown to be exactly those of free space. We then show that simultaneous local-field corrections
to electric {\it and} magnetic fields in purely radiatively broadened magneto-dielectrica have a surprising and potentially
important effect: For sufficiently large densities the real part of the refractive index saturates at the level of $-2$.
At the same time, the imaginary part of the complex index approaches zero inversely proportional to the density.
Thus the medium becomes transparent and left-handed i.e. displays a negative index of refraction with low absorption.


\section{Local-field effects and renormalization of radiative self-interaction in dielectric media}


We start by developing a microscopic scattering approach to local-field effects
in dielectric media taking into account possible material induced modifications of radiative linewidth and
transition frequencies in a self-consistent way. To this end we consider a simple cubic lattice of electric 
point dipoles with isotropic bare polarizability $\alpha_b$ 
\begin{equation}
\alpha_b(\rb)=\alpha_b\sum_\Rb\delta(\rb-\Rb),
\label{alphaE}
\end{equation}
where $\Rb$ denote lattice vectors. The dipoles
interact with the quantized electromagnetic field $\hat \Eb$ which obeys the vector 
Helmholtz equation
\begin{equation}
 {\vec  \nabla}\times{\vec \nabla}\times \hat\Eb(\rb,\omega) -\frac{\omega^2}{c^2}
\hat \Eb(\rb,\omega) =\mu_0\omega^2 \hat \Pb\ .
\label{eq:HelmholtzEDiel}
\end{equation}
In the weak-excitation, i.e. linear response limit, the operator of the microscopic electric polarization
$\hat\Pb$  has the form $\hat \Pb(\rb)=\alpha_b(\rb) \hat\Eb(\rb,\omega)$. Solving
eq.(\ref{eq:HelmholtzEDiel}) we can determine the (isotropic) dispersion relation $k=k(\omega)$ from which
the permittivity $\varepsilon(\omega)$ can be extracted. In the linear response limit the solution of the quantum 
mechanical interaction problem can most easily be obtained by means of Greensfunction techniques. 
In particular it is sufficient to calculate the scattering $T$-matrix of the oscillator lattice.
The dispersion relation can then be obtained via \cite{Korringa,Kohn,Ziman}
\begin{equation}
 {\rm det}\,  T^{-1}=0.
\label{eq:T-matrix1}
\end{equation}
The scattering $T$-matrix  obeys a linear Dyson equation
\begin{equation}
 T = V + V {\cal G}^{(0)} V + \cdots = V + V {\cal G}^{(0)} T,
\label{eq:Dyson}
\end{equation}
where $\Gbn(\rb,\rb^\prime,\omega)$  is the free-space retarded propagator of the electric field
which is a solution to the classical
vector Helmholtz equation  
\begin{eqnarray}
&&{\vec  \nabla}\times{\vec \nabla}\times {\cal G}^{(0)}(\rb,\rb^\prime,\omega) -\frac{\omega^2}{c^2}
{\cal G}^{(0)}(\rb,\rb^\prime,\omega) =\nonumber \\
&&\qquad = \eins\, \delta(\rb-\rb^\prime),
\label{eq:HelmholtzE-GF}
\end{eqnarray}
and 
\begin{equation}
 V(\rb,\omega)=-\frac{\omega^2 \alpha_b(\rb)}{\varepsilon_0 c^2}
\end{equation}
is a linear, isotropic point vertex.
Note that integration over spatial variables was suppressed in eq.(\ref{eq:Dyson}) for notational simplicity.

For a cubic lattice of isotropic scatterers, the series can be summed up to yield \cite{Vries}
\begin{equation}
T(\kb,\kb')= -\!\sum_{\Rb'} e^{i(\kb-\kb')\Rb'}\! \left\{\frac{1}{t(\omega)} +\!\sum_{\Rb\neq 0} e^{i\kb'\Rb}{\cal G}^{(0)}(\Rb)\right\}^{\!\! -1}\!\!\!,
\label{eq:T-matrix}
\end{equation}
where ${\cal G}^{(0)}(\Rb)$ stands for ${\cal G}^{(0)}(\rb,\rb+\Rb,\omega_0)$ which due to the
discrete translation invariance is independent on $\rb$. The single-particle scattering $t$-matrix  $t(\omega)$ is determined 
by the bare polarizability \cite{Lagendijk}
\begin{equation}
t(\omega)^{-1}=\Bigl(\frac{\omega^2}{c^2}\frac{\alpha_b}{\varepsilon_0}\Bigr)^{-1}+\Gb^{(0)}(0).
\label{eq:t-matrix}
\end{equation}
Note that $\Gbn(0)$ is diagonal and isotropic.
In eq.(\ref{eq:T-matrix}) we have separated the contribution of the lattice ($\sum_{\Rb\neq 0}$) from the multiple scattering events at 
the same oscillator ($\Gb^{(0)}(0)$).
This separation is crucial since $\Gb^{(0)}(0)$ is singular.  Rather than eliminating this singularity
by a regularization procedure as done in \cite{Lagendijk}, we note that expression (\ref{eq:t-matrix})
gives the single-particle scattering $t$-matrix $t(\omega)$ dressed by the interaction with the vacuum field
in free space. This quantity is experimentally observable and is related to
the single-particle polarizability $\alpha(\omega)$ in free space:
\begin{equation}
\alpha(\omega)= t(\omega) \, \frac{c^2}{\omega^2}\varepsilon_0
\label{eq:Alpha}
\end{equation}
 $\alpha_b$ on the other hand is not observable and thus only a 
theoretical notion. 
At this point other broadening mechanisms can be incorporated
by adding appropriate non-radiative decay rates $\gamma^\text{non-rad}$ to the
polarizability $\alpha(\omega)$ (\ref{eq:Alpha})
(cf. equation (\ref{eq:alpha}) and discussion thereafter).

Obviously, for the radiative part 
separating the sum 
$\sum_\Rb e^{i\kb'\Rb}\Gbn(\Rb)$ into $\Gbn(0) + \sum_{\Rb\neq 0}e^{i\kb'\Rb}\Gbn(\Rb)$ 
does the trick of writing the full lattice $T$-matrix in terms of the
\emph{known} free space $t$-matrix. As a drawback we are left with 
the sum over the lattice vectors $\Rb\neq 0$. Unfortunately this sum can not be evaluated
exactly and has to be treated approximately.

According to Poisson's summation formula
\begin{equation}
\sum_{n=-\infty}^\infty f(n)=\sum_{k=-\infty}^\infty\int_{-\infty}^\infty \!\!{\rm d}xf(x)e^{-2\pi ikx}
\end{equation}
the sum over $\Rb\neq 0$ can be expressed in terms of a real space integral and a sum over inverse lattice vectors $\Kb$ 
of the Fourier transform of the  free space Greensfunction $\widetilde\Gbn(\pb)$
\begin{equation}
\sum_{\Rb\neq 0} e^{i\kb\Rb}\Gbn(\Rb) = \sum_\Kb\int \!\!{\rm d}\rb {\rm d}\pb \frac{\Xi(|\rb|)}{(2\pi a)^3} e^{i(\pb+\kb-\Kb)\rb}\widetilde \Gbn(p) 
\end{equation}
Here  $\Xi(|\rb|)$ is some smooth function with $\Xi(0)=0$ and $\Xi(|\rb|>0)\to 1$ introduced to prevent the integral from 
touching the excluded singular point $\rb =0$. 

In the following we restrict the discussion to lattices with a lattice constant much smaller than
the resonant wavelength, i.e.  $ka\ll 1$. 
In this limit the lattice of oscillators behaves essentially as a homogeneous medium.
Contributions from large $\Kb$-vectors to the sum, which reflect the discreteness of the lattice, can 
be neglected as long as the singular contribution from the 
origin has been excluded.
Therefore we only keep the term $\Kb=0$ and assume a Gaussian cutting function
$\Xi(|\rb|)=1-e^{-r^2/\delta^2}$, with $\delta \ll a$. This yields
\begin{equation}
\begin{split}
&\sum_{\Rb\neq 0} e^{i\kb\Rb}\Gbn(\Rb) \approx  \frac{1}{a^3}\widetilde\Gbn(k) \\
& \qquad -\frac{1}{a^3}\frac{\pi^{3/2}\delta^3}{(2\pi)^3}\int\!\! {\rm d}\pb\, p^2e^{-\frac{\delta^2}{4}(k^2+p^2)}  e^{-\frac{\delta^2}{2}\kb\cdot \hat\pb}\widetilde\Gbn(p),
\end{split}
\end{equation}
where $\hat \pb = \pb/|\pb|$.
Apart from the Gaussian $p$-integral which provides a smooth cut-off in reciprocal space,
$\delta$ can be treated as a small parameter. That allows to carry out the integration
analytically which in leading order of $\delta$ yields
\begin{equation}
\begin{split}
\sum_{\Rb\neq 0} e^{i\kb\Rb}\Gbn(\Rb) \approx & \frac{1}{a^3}\widetilde\Gbn(k)-\frac{1}{a^3}\frac{1}{3\omega^2/c^2}\eins.
\end{split}
\end{equation}
The free-space Greentensor $\widetilde\Gbn(k)$ is given by \cite{Lagendijk}
\begin{equation}
\widetilde\Gbn(k) = \left(\frac{\omega^2}{c^2}\eins- |\kb|^2\Delta_{\kb}\right)^{-1}
\end{equation}
with $\Delta_{\kb}=\eins-\hat\kb\otimes\hat\kb$ being a projector to directions orthogonal to
$\kb$.

With this we are ready to evaluate eq. (\ref{eq:T-matrix1}) which reads in the limit $ka\ll 1$
\begin{equation}
\det \left(\frac{1}{\frac{\omega^2}{c^2}\rho \alpha(\omega)/\varepsilon_0}\eins +\frac{1}{\frac{\omega^2}{c^2}\eins- |\kb|^2\Delta_{\kb}}
-\frac{\eins}{3\frac{\omega^2}{c^2}}\right)=0.
\label{det}
\end{equation}
Solving eq. (\ref{det}) for the (isotropic) dispersion $k=k(\omega)$ with $k(\omega)=\varepsilon(\omega)\omega^2/c^2$ 
finally yields
\begin{equation}
\varepsilon(\omega)=1+\frac{\rho\, \alpha(\omega)/\varepsilon_0}{1-\rho\, \alpha(\omega)/3\varepsilon_0}.
\end{equation}
This is the well-known Clausius-Mossotti relation where for purely radiatively broadened systems
$\alpha(\omega)$ is the dressed polarizability
of an isolated oscillator interacting with the free-space electromagnetic vacuum field.


\section{Local-field effects for magneto-dielectrics}


We now extend the above discussion to the case of a bi-cubic lattice of electric and magnetic 
dipole oscillators.
The microscopic, space-dependent bare electric polarizability $\alpha_{be}(\rb)$ 
is then given by
\begin{equation}
\alpha_{be}(\rb)=\alpha_{be}\sum_\Rb\delta(\rb-\Rb)=\frac{\alpha_{be}}{a^3}\sum_\Kb e^{i\Kb\rb}
\label{alphae}
\end{equation}
and, similarly, the bare magnetic polarizability by
\begin{equation}
\alpha_{bm}(\rb)=\alpha_{bm}\sum_\Rb\delta(\rb-\Rb-\Delta\rb)=\frac{\alpha_{bm}}{a^3}\sum_\Kb e^{i\Kb(\rb-\Delta\rb)}
\label{alpham}
\end{equation}
Here $\Rb$ denotes again the lattice vectors  and $\Delta\rb$ the spacing between the electric and magnetic sublattices. 
The bare atomic polarizabilities $\alpha_{be}$ and $\alpha_{bm}$ are assumed to be scalar for simplicity corresponding
to an isotropic medium.
The last expressions in eqn. (\ref{alphae}) and (\ref{alpham}) give the bare polarizabilities
in reciprocal space, with $\Kb$ being the reciprocal lattice vectors.

Due to the simultaneous presence of electric and magnetic dipole lattices we now have to solve the coupled
set of vector Helmholtz equations for the operators of the electric and magnetic fields
\begin{equation}
\nabla\times\nabla\times\hat \Eb -\frac{\omega^2}{c^2}\hat \Eb=i\omega\mu_0\nabla\times\hat \Mb+\mu_0\omega^2\hat \Pb
\label{HelmholtzE}
\end{equation}
and 
\begin{equation}
\nabla\times\nabla\times\hat \Hb -\frac{\omega^2}{c^2}\hat\Hb=\frac{\omega^2}{c^2}\hat\Mb-i\omega\nabla\times\hat\Pb.
\label{HelmholtzH}
\end{equation}
In linear response the operator of the polarization $\hat\Pb$ and the
magnetization $\hat \Mb$ are  proportional to the electric and magnetic fields
respectively, $\hat\Pb(\rb)=\alpha_{be}(\rb)\hat\Eb(\rb)$ 
and $\hat\Mb(\rb)=\mu_0\alpha_{bm}(\rb)\hat\Hb(\rb)$.

In the following we will pursue a slightly different approach to solve the coupled set of equations
than used in the previous section. 
Taking into account the lattice symmetry we first write the field variables in the form
\begin{equation}
\hat\Eb(\rb)=\int_{\rm 1.BZ}\!\!{\rm d}\kb\, \sum_\Kb \widetilde\Eb(\kb-\Kb)e^{i(\kb-\Kb)\rb},
\label{fieldFourier}
\end{equation}
where the dependence on frequency $\omega$ was suppressed for
notational simplicity. The subscript denotes integration over the first Brillouin zone.
Substituting this and the corresponding expression for $\hat\Hb$
 into (\ref{HelmholtzE})-(\ref{HelmholtzH})
gives the Helmholtz equations in reciprocal space.
After some elementary manipulations the following closed set of equations is derived:
\begin{widetext}
\begin{equation}
\left[\frac{\eins}{\frac{\omega^2}{c^2}\rho\alpha_{be}/\varepsilon_0}+\sum_\Kb\frac{1}{\frac{\omega^2}{c^2}\eins
-|\kb-\Kb|^2\Delta_{\kb-\Kb}}\right]\sum_{\Kb'}\widetilde\Eb(\kb-\Kb') = \frac{\mu_0\alpha_{bm}}{\omega\alpha_{be}} \sum_\Kb\frac{e^{i\Kb\Delta\rb}(\kb-\Kb)\times}{\frac{\omega^2}{c^2}\eins-|\kb-\Kb|^2\Delta_{\kb-\Kb}}
\sum_{\Kb'}\widetilde\Hb(\kb-\Kb')e^{-i\Kb'\Delta\rb}
\label{HelmholtzReciprocalE}
\end{equation}
\begin{equation}
\left[\frac{\eins}{\frac{\omega^2}{c^2}\rho\mu_0\alpha_{bm}}+\sum_\Kb\frac{1}{\frac{\omega^2}{c^2}\eins
-|\kb-\Kb|^2\Delta_{\kb-\Kb}}\right]\sum_{\Kb'}\widetilde\Hb(\kb-\Kb')e^{-i\Kb'\Delta\rb} = -\frac{c^2\alpha_{be}}
{\omega\mu_0\alpha_{bm}} \sum_\Kb\frac{e^{-i\Kb\Delta\rb}(\kb-\Kb)\times}{\frac{\omega^2}{c^2}\eins-|\kb-\Kb|^2\Delta_{\kb-\Kb}}\sum_{\Kb'}\widetilde\Eb(\kb-\Kb')
\label{HelmholtzReciprocalH}
\end{equation}
\end{widetext}
where $\rho=1/a^3$ is the particle density.
The sum in the brackets on the left hand sides of eqs. (\ref{HelmholtzReciprocalE},\ref{HelmholtzReciprocalH})
can be rewritten as
\begin{eqnarray*}
&&\rho \sum_\Kb\frac{1}{\frac{\omega^2}{c^2}{\mathbbm 1}-|\kb-\Kb|^2\Delta_{\kb-\Kb}}
= \sum_{\Rb} e^{i\kb\Rb}{\cal G}^{(0)}(\Rb) \\
&&\qquad \qquad= \Gbn(0) + \sum_{\Rb\neq 0} e^{i\kb\Rb}{\cal G}^{(0)}(\Rb), 
\end{eqnarray*}
where in the second line we have separated the singular contribution $\Gbn(0)$. One recognizes
that this term can be added to the expressions containing the bare polarizabilities in eqs.(\ref{HelmholtzReciprocalE}) and
(\ref{HelmholtzReciprocalH}) yielding the dressed scattering $t$-matrices for isolated electric and magnetic dipoles
interacting with the free-space vacuum field:
\begin{eqnarray}
t_e(\omega)^{-1} &=&\Bigl(\frac{\omega^2}{c^2}\frac{\alpha_{be}}{\varepsilon_0}\Bigr)^{-1}+\Gb^{(0)}(0),\\
t_m(\omega)^{-1}& =&\Bigl(\frac{\omega^2\mu_0}{c^2}\alpha_{bm}\Bigr)^{-1}+\Gb^{(0)}(0).
\end{eqnarray}
The sum over the Greensfunction excluding $\Rb=0$ can be evaluated in a similar way as in the previous
section. If we again assume a lattice constant $a$ much smaller than the resonant wavelength, reciprocal
$\Kb$ vectors different from zero can be disregarded. This leads to
\begin{eqnarray}
&&\left[\frac{1}{\rho t_e(\omega)} +\widetilde\Gbn(\kb) -\frac{1}{3\omega^2/c^2}\right]\hat \Eb(\kb) = \label{HelmholtzReciprocalE-short}\\
&& \qquad \qquad=\frac{\mu _0 \alpha_{bm}}{\omega \alpha_{be}} \frac{\kb \times}{\frac{\omega^2}{c^2}\eins - k^2\Delta_{\kb}}
\hat \Hb(\kb),\nonumber\\
&&\left[\frac{1}{\rho t_m(\omega)} +\widetilde\Gbn(\kb) -\frac{1}{3\omega^2/c^2}\right]\hat \Hb(\kb) = \label{HelmholtzReciprocalH-short}\\
&& \qquad \qquad=\frac{c^2 \alpha_{be}}{\omega \mu_0\alpha_{bm}} \frac{\kb \times}{\frac{\omega^2}{c^2}\eins - k^2\Delta_{\kb}}
\hat \Eb(\kb).\nonumber
\end{eqnarray}
Since we are furthermore only interested in propagating, i.e. transversal modes, we can
further simplify the calculation by projecting onto transversal modes using $\Delta_{\kb}$
\begin{widetext}
\begin{equation}
\left[\frac{1}{\frac{\omega^2}{c^2}\rho\alpha_e(\omega)/\varepsilon_0}
+\frac{1}{\frac{\omega^2}{c^2}-k^2}-\frac{1}{3\frac{\omega^2}{c^2}}\right]\Delta_\kb
\hat\Eb(\kb) = \frac{\mu_0\alpha_{bm}}{\omega\alpha_{be}} \frac{1}{\frac{\omega^2}{c^2}-k^2}\kb\times\Delta_\kb\hat\Hb(\kb)
\label{HelmholtzReciprocalErenorm}
\end{equation}
\begin{equation}
\left[\frac{1}{\frac{\omega^2}{c^2}\rho\mu_0\alpha_m(\omega)}
+\frac{1}{\frac{\omega^2}{c^2}-k^2}-\frac{1}{3\frac{\omega^2}{c^2}}\right]\kb\times\Delta_\kb
\hat\Hb(\kb)= \frac{c^2\alpha_{be}}{\omega\mu_0\alpha_{bm}} \frac{k^2}{\frac{\omega^2}{c^2}-k^2}\Delta_\kb
\hat\Eb(\kb).
\label{HelmholtzReciprocalHrenorm}
\end{equation}
\end{widetext}

Here we have substituted the dressed single particle $t$-matrices by the free-space dressed polarizabilities
$\alpha_{e(m)}(\omega)= t_{e(m)}(\omega) c^2/\omega^2\varepsilon_0\left(\mu_0^{-1}\right)$.

In order to find the dispersion $k(\omega)=n^2\omega^2/c^2$ we 
have to determine the solution of the secular equation of 
the linear set of eqs. (\ref{HelmholtzReciprocalErenorm},\ref{HelmholtzReciprocalHrenorm}), which results
in the condition
\begin{eqnarray}
&& \left[\frac{1}{\frac{\omega^2}{c^2}\rho\alpha_e(\omega)/\varepsilon_0}
+\frac{1}{\frac{\omega^2}{c^2}-k^2}-\frac{1}{3\frac{\omega^2}{c^2}}\right]\times \\
&&\qquad\times \left[\frac{1}{\frac{\omega^2}{c^2}\rho\mu_0\alpha_m(\omega)}
+\frac{1}{\frac{\omega^2}{c^2}-k^2}-\frac{1}{3\frac{\omega^2}{c^2}}\right]=0.\nonumber
\end{eqnarray}
Solving for the refractive index of the transversal modes then gives $n^2=\varepsilon\mu$,
where
\begin{equation}
\varepsilon=1+\frac{\rho\alpha_e(\omega)/\varepsilon_0}{1-\rho\alpha_e(\omega)/3\varepsilon_0}
\end{equation}
\begin{equation}
\mu=1+\frac{\rho\mu_0\alpha_m(\omega)}{1-\rho\mu_0\alpha_m(\omega)/3}
\end{equation}
are the relative dielectric permittivity and magnetic permeability, respectively, both satisfying
the Clausius-Mossotti relations.

Note that for longitudinal modes eqs. (\ref{HelmholtzReciprocalE-short}) and (\ref{HelmholtzReciprocalH-short}) decouple.
This can be seen by applying the corresponding projector to longitudinal waves $\hat\kb\otimes\hat\kb$ which leads to a disappearance of the cross-coupling terms.
The dispersion obtained in this way gives either $\varepsilon=0$ corresponding to electric excitons
\cite{Davydov,Agranovich} or $\mu=0$ for magnetic excitons.


\section{negative refraction and absorption reduction due to local field
effects in magneto-dielectric media}


It is interesting to consider the implications of the Clausius Mossotti relations for radiatively broadened
media in the large density limit. Let us first consider a purely dielectric medium
and let us assume that the polarizability $\alpha_e(\omega)=\alpha_e^\prime(\omega) + i\, \alpha_e^{\prime\prime}(\omega)$
 does not depend on the density, i.e. the medium is radiatively broadened. In this case
one finds 
\begin{equation}
\varepsilon(\omega) \enspace \stackrel{\rho\to \infty}{\longrightarrow}\enspace  -2 + i\frac{1}{\rho}\frac{9\varepsilon_0 \alpha_e^{\prime\prime}}{|\alpha_e|^2}.
\end{equation}
In the high-density limit and  sufficiently close to resonance the response saturates
at a value of $-2$ with an imaginary part that vanishes as $1/\rho$.
At this point the medium becomes totally opaque since the index of refraction attains
an imaginary value $n=i\sqrt{2}$ indicating the emergence of a stopping band. 
%
%
\begin{figure}[ht]
\epsfig{file=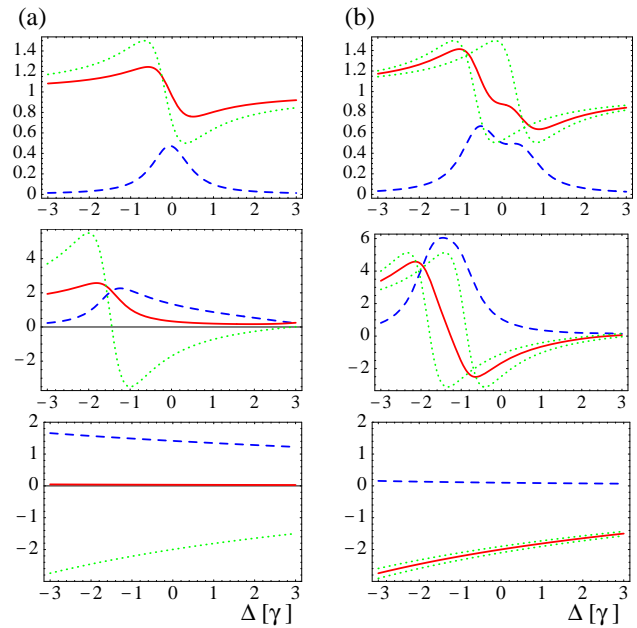,width=8.5 cm}
\caption{(color online) spectrum of the real ({\color{red} solid}) and 
imaginary ({\color{blue} dashed}) part of the refractive index as well as the real ({\color{green} dotted})
part of the response function(s) $\varepsilon$ and/or $\mu$ as a function of the detuning $\Delta$ for a
(a) pure dielectric or magnetic medium for $\rho\abs{\alpha_0}/3$ at $\Delta=0$ equal to $=1/3$ (top), $3$ (middle)
and $30$ (bottom)
(b) magneto-dielectric medium for $\rho\abs{\alpha_0}/3$ at $\Delta=0$ equal to $=1/3$ (top), $3$ (middle)
and $30$  (bottom).
}
\label{fig1}
\end{figure}
%
%
This is illustrated in the left column of Fig \ref{fig1} for a medium composed of either electric or magnetic dipole
oscillators. For small densities ($\rho\abs{\alpha_0}/3=1/3$) the resonance is centered at $\omega_0$ whereas 
for larger densities ($\rho\abs{\alpha_0}/3=3$) the response shifts to smaller frequencies and is amplified. 
Eventually ($\rho\abs{\alpha_0}/3=30$) the refractive index becomes almost purely imaginary in which case 
light cannot propagate any longer.

This behavior changes dramatically if we consider media with overlapping  electric and magnetic resonances
described by both an electric polarizability $\alpha_e(\omega)$ 
and a magnetic polarizability $\alpha_m(\omega)$. 
Independent application of Clausius-Mossotti local-field corrections to the permittivity and the
permeability  leads in the high density limit to
\begin{equation}
n=-2+i\frac{1}{\rho}\left( \frac{9 \varepsilon_0 \alpha_e^{\prime\prime}}{\abs{\alpha_e}^2}+\frac{9 \alpha_m^{\prime\prime}}{\mu_0 \abs{\alpha_m}^2} \right).
\end{equation}
Thus in the spectral overlap region the real part of the index of refraction approaches the value $-2$, i.e. attains a
constant negative value. Furthermore the imaginary part, responsible for absorption losses, approaches zero in that spectral region as $1/\rho$. This rather peculiar behavior is illustrated in the right column of Fig.\ref{fig1}. One clearly recognizes the emergence of a spectral region around the bare resonance frequency where the real part of the refractive index approaches $-2$ 
while the imaginary part is strongly suppressed. 

Negative refraction of light
is currently one of the most active research areas in photonics 
\cite{Veselago68,AgranovichRef,Shalaev} due to fascinating potential applications such as superlensing \cite{Pendry00}
or electromagnetic cloaking \cite{Leonhardt,Pendry06,Schurig}. In recent years 
substantial progress has been made in realizing 
negative refraction in so-called meta-materials \cite{Pendry99,Smith00,Yen04,Linden04}. 
These are artificial periodic structures of electric and magnetic dipoles 
with a resonance wavelength much larger than the lattice constant which thus 
form a quasi-homogeneous magneto-dielectric medium. In order to achieve a
large electromagnetic response, operation close to
resonance is needed which is associated with rather substantial losses. 
The elimination of these losses represents one of the main challenges in the
field \cite{Kaestel-submitted}. We have shown here that in a radiatively broadened medium, i.e.
a medium in which density-dependent broadening mechanism can still be disregarded
for sufficiently large densities, local field effects can provide a negative index of refraction and
at the same time efficiently suppress absorption losses.


\section{summary}


In the present paper we have given a rigorous microscopic derivation of Clausius-Mossotti relations for both the electric
and magnetic response in an isotropic, radiatively broadened magneto-dielectric medium formed by a simple bi-cubic lattice
of electric and magnetic dipoles. As opposed to previous microscopic approaches we have taken into account possible
modifications of the single-particle polarizabilities by the altered electromagnetic vacuum inside the medium in
a self-consistent way. For a simple bi-cubic lattice it has been shown that the polarizabilities entering the 
Clausius-Mossotti relations are those of single oscillators interacting with the free-space vacuum field.
We showed that as a consequence of the local field corrections 
a radiatively broadened medium with overlapping electric and magnetic resonances
becomes lossless with a real part of the refractive index approaching the value $-2$ in the 
high-density limit.  The latter could provide an interesting avenue to construct artificial materials
with negative refraction and low losses.


\

\

\begin{acknowledgments}
This work was supported by the Alexander von Humboldt Foundation through the 
institutional collaboration grant between The Institute of Theoretical Physics and Astronomy
of Vilnius University and the Technical University of Kaiserslautern.
J.K. acknowledges financial support by the Deutsche Forschungsgemeinschaft
through the GRK 792 ``Nichtlineare Optik und Ultrakurzzeitphysik''.
\end{acknowledgments}

\def\etal{\textit{et al.~}}

\end{document}